\documentclass[twocolumn]{aastex63}

\shorttitle{The BoRG-\jwst\ Galaxies as Analogs of the $z\gtrsim10$ Blue Monsters}
\shortauthors{Rojas-Ruiz et al.}

\usepackage{float}
\usepackage{natbib}
\usepackage{hyperref}
\usepackage[nointegrals]{wasysym}
\usepackage{ragged2e}
\usepackage{graphicx}
\usepackage{units}
\usepackage{amsmath}
\definecolor{red}{rgb}{1,0,0}
\definecolor{orange}{RGB}{204, 85, 0}
\definecolor{blue}{HTML}{4169e1}
\definecolor{ltred}{RGB}{245,167,162}
\definecolor{ltblue}{RGB}{206,211,242}

\newcommand{\heii}{He\,{\sc ii}}
\newcommand{\ciii}{C\,{\sc iii}]}
\newcommand{\civ}{C\,{\sc iv}}

\newcommand{\oiii}{[O\,{\sc iii}]}
\newcommand{\oiiS}{[O\,{\sc ii}]}

\newcommand{\neiii}{Ne\,{\sc iii}}

\newcommand{\Msun}{M$_\odot$}

\newcommand{\hst}{\textit{HST}}
\newcommand{\jwst}{\textit{JWST}}

\defcitealias{roberts-borsani_borg-jwst_2025}{RB25}
\defcitealias{rojas-ruiz_probing_2020}{RR20}
\defcitealias{leethochawalit_uv_2023}{L23}
\defcitealias{rojas-ruiz_borg-jwst_2025}{RR25}
\begin{document}

\title{The BoRG-\textit{JWST} Survey: Analogs at $z\sim8$ to the UV-luminous Galaxy Population at $z\gtrsim10$}

\correspondingauthor{Sof\'ia Rojas-Ruiz}
\email{rojas@astro.ucla.edu}

\author[0000-0003-2349-9310]{Sof\'ia Rojas-Ruiz}\affiliation{Department of Physics and Astronomy, University of California, Los Angeles, 430 Portola Plaza, Los Angeles, CA 90095, USA}

\author[0000-0002-4140-1367]{Guido Roberts-Borsani}
 \affiliation{Department of Physics \& Astronomy, University College London, London, WC1E 6BT, UK}

\author[0000-0002-8512-1404]{Takahiro Morishita}
\affiliation{IPAC, California Institute of Technology, MC 314-6, 1200 E. California Boulevard, Pasadena, CA 91125, USA}

\author[0000-0003-2536-1614]{Antonello Calabrò}
\affiliation{INAF--Osservatorio Astronomico di Roma, via di Frascati 33, 00078 Monte Porzio Catone, Italy}

\author[0000-0002-9921-9218]{Micaela Bagley}
\affiliation{Department of Astronomy, The University of Texas at Austin, 2515 Speedway, Austin, TX, 78712, USA}

\author[0000-0002-8460-0390]{Tommaso Treu}
\affiliation{Department of Physics and Astronomy, University of California, Los Angeles, 430 Portola Plaza, Los Angeles, CA 90095, USA} 

\author[0000-0001-8519-1130]{Steven L. Finkelstein}
\affiliation{Department of Astronomy, The University of Texas at Austin, 2515 Speedway, Austin, TX, 78712, USA}

\author[0000-0001-9935-6047]{Massimo Stiavelli}
\affiliation{Space Telescope Science Institute, 3700 San Martin Drive, Baltimore, MD 21218, USA}
\affiliation{Dept. of Physics \& Astronomy, Johns Hopkins University, Baltimore, MD 21218, USA}
\affiliation{Dept. of Astronomy, University of Maryland, College Park, MD 20742, USA}

\author[0000-0001-9391-305X]{Michele Trenti}
\affiliation{School of Physics, University of Melbourne, Parkville 3010, VIC, Australia}

\author[0000-0003-3466-035X]{{L. Y. Aaron} {Yung}}
\affiliation{Space Telescope Science Institute, 3700 San Martin Drive, Baltimore, MD 21218, USA}

%=====================================================
%============ Abstract ===============================
%====================================================
\begin{abstract}
The population of bright galaxies at $z\gtrsim10$ discovered by JWST, including the so-called ``blue monsters”, has been difficult to reconcile with standard galaxy evolution models. To shed light on this extraordinary population, we study the $z\sim8$ galaxies discovered by the BoRG-\jwst\ survey. These slightly-lower redshift analogs are comparable in UV luminosity to the blue monsters, and their lower redshift makes it much easier to access key rest frame optical diagnostics with NIRspec.   
We find that BoRG-\jwst\ galaxies are consistent with being dust-poor based on their blue UV slopes and Balmer decrement ratios. We find no strong evidence for dominant active galactic nuclei contribution to the UV brightness, based on line-ratio diagnostics, though some contribution cannot be excluded.
We further infer the stellar mass, star formation and UV-brightness history of the BoRG-\jwst\ galaxies by fitting their rest-frame UV-optical spectra.
We see evidence for stochastic episodes of star formation for all the BoRG-\jwst\ galaxies, providing a temporal boosting of UV luminosity in short timescales. The UV-bright blue monsters at $z\gtrsim10$ can be explained by the presence of stars with ages below 100~Myr.

\end{abstract}


\section{Introduction}\label{intro}
It took just a few days for the \jwst\ Near-Infrared Camera (NIRcam; 0.6--5$\mu$m) to force a revision on our understanding of early galaxy formation by discovering numerous candidate bright galaxies at $z\gtrsim10$ \citep[e.g.,][]{castellano_early_2022,finkelstein_long_2022,naidu_two_2022}. Prior to \jwst, surveys robust against cosmic variance were already sensitive to rare, intrinsically luminous galaxies at $z\sim6-9$, including wide-are ground-based surveys \citep[e.g.,][]{stefanon_brightest_2019,bowler_lack_2020}, and Hubble Space Telescope (\hst) random pointings \citep[e.g.,][]{trenti_brightest_2011,bradley_brightest_2012,bernard_galaxy_2016,calvi_bright_2016,morishita_bright-end_2018,rojas-ruiz_probing_2020,leethochawalit_uv_2023,bagley_bright_2024}. These studies reported a higher-than-expected abundance of bright systems relative to simple extrapolations from fainter populations, foreshadowing the tensions later amplified by the first discoveries of UV-luminous galaxies at $z>10$ with \jwst. Multiple \jwst\ studies have since found more of these galaxies \citep[e.g.,][]{castellano_early_2023,curtis-lake_spectroscopic_2023,finkelstein_ceers_2023,finkelstein_complete_2024,harikane_comprehensive_2023,harikane_jwstnirspec_2023,robertson_earliest_2024,hainline_cosmos_2024}, which has enabled detailed characterization of this population. For example, these galaxies have been found to have very blue UV-continuum slopes ($\beta_{\mathrm{UV}}$), with indications of ``dust-free'' or essentially no dust attenuation for $z\gtrsim11$ galaxies \citep[e.g.,][]{austin_large_2023,cullen_ultraviolet_2024,morales_rest-frame_2024, topping_uv_2024}. Additionally, the population of $z>10$ galaxies with intrinsically bright ultraviolet (UV) luminosities ($M_{\rm{UV}} \lesssim -20$)  appears to have intermediate to high stellar masses of $\sim10^8-10^9$\Msun \citep[e.g.,][]{bunker_jades_2023,carniani_spectroscopic_2024,castellano_jwst_2024,naidu_cosmic_2025}, with extreme cases pushing the limits of early mass assembly \citep[e.g.,][]{casey_cosmos-web_2024}. This population of massive UV-bright galaxies has been aptly nicknamed ``blue monsters'' to describe their extreme properties \citep[e.g.,][]{ziparo_blue_2023,ferrara_eventful_2024,ferrara_blue_2025}. 

Three broad classes of mechanisms have been proposed to explain the extreme properties observed in blue monsters: (i) The lack of dust, which could be explained by ejection due to radiation pressure or segregated with respect to UV-emitting regions \citep{ferrara_stunning_2023,ferrara_blue_2025,ziparo_blue_2023}. Some studies have already favored the dust ejection case given non-detection of dust continuum at sensitive wavelengths tracing dust emission in these UV-luminous galaxies \citep{bakx_deep_2023,fujimoto_alma_2023,carniani_eventful_2025}. (ii) The nature of the source and mechanisms responsible for removing dust from these early galaxies. \citet{fiore_dusty-wind-clear_2023} proposed radiative outflows clearing dust in the galaxy that could be produced by the presence of active galactic nuclei (AGN) \citep{bunker_jades_2023,maiolino_author_2024,maiolino_small_2024}, low-metallicity stellar populations \citep{ferrara_eventful_2024}, or dual AGN and strong star formation contributing to the emission \citep{calabro_evidence_2024,napolitano_dual_2025}. (iii) Stochastic star formation and thus young stellar ages at the time of observations can explain the extreme luminosity \citep{mason_brightest_2023,casey_cosmos-web_2024,dressler_early_2023,dressler_building_2024,gelli_impact_2024}. Young stellar ages alone can produce elevated light-to-mass ratios \citep[e.g.,][]{donnan_very_2025}, and burstiness in star formation may naturally arise if early galaxies host higher-density gas reservoirs, which in turn allow more strongly modulated and higher star-formation efficiencies at these redshifts \citep{finkelstein_complete_2024,somerville_density_2025}.

Importantly, however, \jwst-NIRSpec observations (0.6–5.3$\mu$m) which have spectroscopically confirmed blue monsters at $z>10$ only probe their rest-frame UV emission. To further investigate the formation and evolution of these UV-luminous galaxies, access to their rest-frame optical emission is necessary \citep[i.e., with MIRI $>5\mu$m][]{ferrara_eventful_2024}. In fact, thus far, just three of these blue monsters have been followed up with MIRI to analyze the \oiii\,$\lambda \lambda$4959,5007, and H$\alpha$ emission lines \citep[MACS0647‑JD; z=10.17,][]{hsiao_jwst_2024}, \citep[GN-z11; z=10.60,][]{alvarez-marquez_insight_2025}, \citep[GHZ2; z=12.33,][]{zavala_luminous_2025}, \citep[JADES-GS-z14-0; z=14.18,][]{helton_ionizing_2025}. The difficult and time-consuming process of detecting the rest-optical emission at $z\gtrsim10$ clearly exemplifies why it is very helpful to study lower redshift analogs.

The BoRG-\jwst\ survey --Cy1 GO 1747 (PI: Roberts-Borsani) and GO 2426 (Co-PIs: Bagley and Rojas-Ruiz)--  provides a sample of spectroscopically confirmed $z=7-9$ galaxies that is robust against cosmic variance \citep{roberts-borsani_borg-jwst_2025}. The NIRSpec PRISM R $=3-300$ resolution observations offer crucial information on the properties of these galaxies in the rest-UV and in the rest-optical spectrum. \cite{roberts-borsani_borg-jwst_2025} calculated the redshifts, $\beta_{\rm{UV}}$ slopes and $M_{\rm{UV}}$ values of the BoRG-\jwst\ galaxies at $z\sim8$ and showed that they are in good agreement with the spectroscopically confirmed luminous $z>10$ population. Furthermore, \cite{rojas-ruiz_borg-jwst_2025} used this sample to measure the UV luminosity function at $z=7-9$, confirming a higher number density of galaxies relative to pre- and early-\jwst\ expectations, alleviating tensions raised by the initially reported 
$z>10$ overabundance, and supporting a stochastic star-formation scenario, as further indicated by the scatter in their mass-to-light ratios. Thus, the BoRG-\jwst\ galaxies are indicative of being ideal analogs to the $z\gtrsim10$ blue monster population based on their intrinsic UV properties. 

In this paper, we use BoRG-\jwst\ galaxy spectra as analogs to the $z>10$ blue monsters that have been spectroscopically confirmed to investigate the physical mechanisms driving the formation of UV-bright, massive galaxies. We describe the data and fitting methodologies to measure emission lines and model star formation histories in \S \ref{methods}. We use the $\beta_{\mathrm{UV}}$ and Balmer decrement H$\delta$/H$\beta$ diagnostics to investigate dust extinction in \S \ref{lack-dust}. We present the OHNO diagnostic to evaluate the AGN vs. strong star-forming engine responsible for the extreme UV emission observed in \S \ref{agn-sf}. Lastly, in \S \ref{bursts} we present their stellar mass, star formation history and UV luminosity as a function of lookback time to identify bursty episodes of star formation. In \S \ref{conclusion} we summarize our findings and highlight the rest-optical properties of BoRG-\jwst\ galaxies that are key to understand blue monsters.

Throughout this work, we use the cosmology according to H$_0$= 70 km~s$^{-1}$ Mpc$^{-1}$, $\Omega_{\rm m}$= 0.3, and $\Omega_\Lambda$ = 0.7 and a \citet{chabrier_galactic_2003} initial mass function (IMF). All magnitudes are given in the AB system.

\section{Data and Methods}\label{methods}

The spectral information offers a holistic assessment of the dust content, ionizing nature, and star formation history responsible for producing these massive UV-bright galaxies. We use the reduced NIRSpec spectra of the BoRG-\jwst\ galaxies as described by \citet{roberts-borsani_borg-jwst_2025}, hereafter \citetalias{roberts-borsani_borg-jwst_2025}. The analysis is divided into emission line diagnostics and modeling of their spectra.

\begin{figure*}[ht!]
\centering
\includegraphics[width=0.325\linewidth]{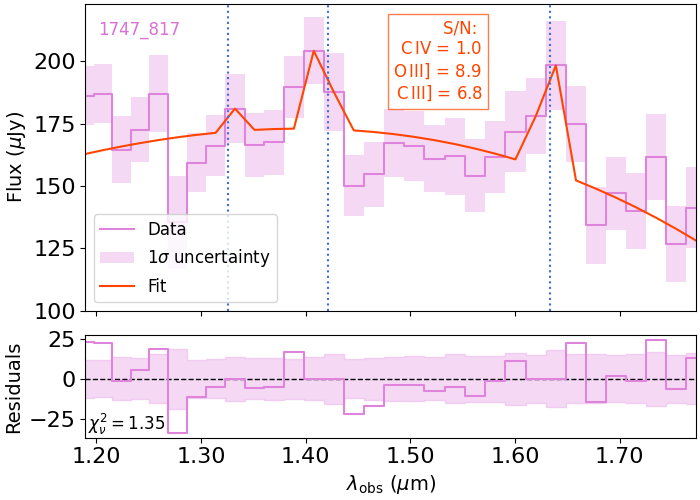}
\includegraphics[width=0.325\linewidth]{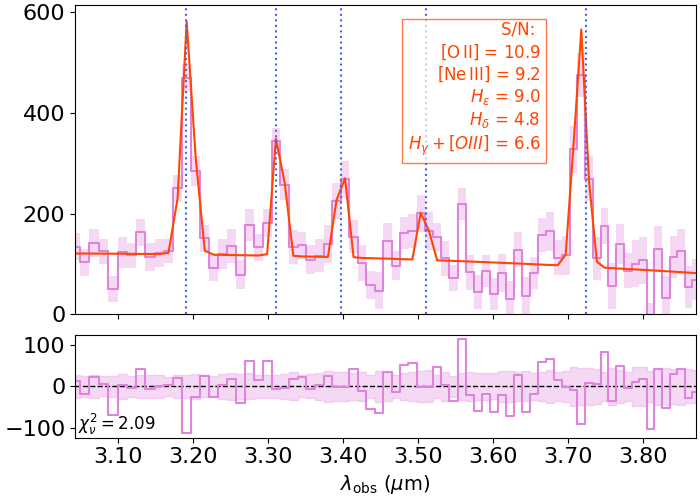}
\includegraphics[width=0.325\linewidth]{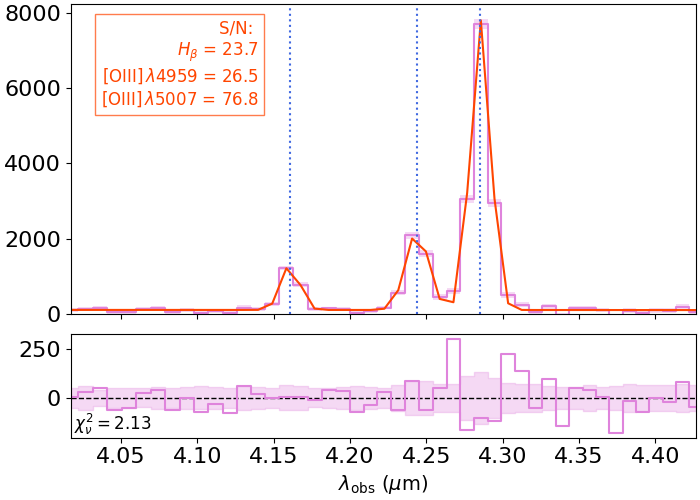} 
\caption{Example of galaxy 1747\_817 showing the emission line fitting in the three spectral windows used in this work. The purple lines and shaded regions indicate the observed flux and 1$\sigma$ uncertainties, while the orangered lines show the Gaussian fits described in \S\ref{emission-fit}. Signal-to-noise ratios for the emission lines are listed ordered left to right as they appear in the spectrum.}
\label{fig:line-fits}
\end{figure*}

\subsection{Emission Line fit}\label{emission-fit}
We divide the spectrum of each galaxy into three rest-frame spectral windows: (1) the rest-UV covering emission lines \civ\,$\lambda \lambda$1548,1550 to \ciii\,$\lambda \lambda$1907,1909; (2) rest-optical spanning \oiiS\,$\lambda \lambda$3726,3729 to H$\gamma$+\oiii\,$\lambda$4363; and (3) a second rest-optical window covering H$\beta$\,$\lambda$4861 through \oiii\,$\lambda \lambda$4959,5007. We use the \texttt{astropy.modeling} package to fit the emission lines using Gaussian profiles. We perform bootstrap resampling 100 times where in each step the spectrum is perturbed with Gaussian noise.  The code iterates over each spectral window on the same noisy realization of the spectrum and computes the line fluxes with the underlying continuum. The minimum signal-to-noise ratio (S/N) required for emission line detection is S/N $ > 2.5$. We also measure the line ratio of H$\delta$/H$\beta$ that is used for dust attenuation analysis in \S \ref{lack-dust}, and the ratios \neiii\,$\lambda$3870/\oiiS\,$\lambda \lambda$3726,3729 and \oiii\,$\lambda$5007/H$\beta$ to identify the type of ionizing source  in \S \ref{agn-sf}. The uncertainties in the emission line fluxes and ratios are calculated from the 16th and 84th percentiles of the bootstrapped distributions. See Figure~\ref{fig:line-fits} for an example of the emission line fits in the three spectral windows for galaxy 1747\_817, and Table~\ref{tabla} for the best-fit parameters of each BoRG-\jwst\ galaxy.

\begin{deluxetable*}{llcccccccc}
\tablecaption{Properties of the BoRG-JWST galaxies}
\tablewidth{1000pt}
\tablehead{
\colhead{ID} &
\colhead{$z$} &
\colhead{M$_{UV}$ } &
\colhead{$\beta_{UV}$ } &
\colhead{\oiiS\,$\lambda \lambda$3726,29} &
\colhead{\oiii\,$\lambda$5007} &
\colhead{H$_\delta$/H$_\beta$} &
\colhead{Ne3O2} &
\colhead{R3} \\
\colhead{} & \colhead{} &\colhead{} & \colhead{} &
\colhead{10$^{-18}$ [cgs]} & \colhead{10$^{-18}$ [cgs]} & \colhead{} & \colhead{} & \colhead{} 
}
\startdata
1747\_138 & 7.179 & $-$21.42$\pm$0.16 & $-2.52 \pm 0.05$  & 1.03$_{-0.17}^{+0.21}$ & 7.45$_{-0.31}^{+0.28}$ & 0.33$_{-0.19}^{+0.10}$ & $-0.12_{-0.18}^{+0.44}$ & 0.60$_{-0.05}^{+0.07}$\\
2426\_112 & 7.337 & $-$21.94$\pm$0.07 & $-2.41 \pm 0.07$ & 1.35$_{-0.19}^{+0.22}$ & 19.51$_{-0.87}^{+0.90}$ & 0.27$_{-0.12}^{+0.12}$ & 0.41$_{-0.18}^{+0.13}$ & 0.82$_{-0.06}^{+0.08}$\\
1747\_1425 & 7.553 &  $-$21.38$\pm$0.06  &  $-2.30 \pm 0.07$ & 1.36$_{-0.29}^{+1.09}$ & 16.29$_{-0.34}^{+0.35}$ & --- & 0.13$_{-0.12}^{+0.24}$ & 0.85$_{-0.04}^{+0.13}$\\
1747\_817 & 7.556 &  $-$20.74$\pm$0.13 & $-2.42 \pm 0.09$ & 2.69$_{-0.25}^{+0.26}$ & 20.59$_{-0.27}^{+0.37}$  & 0.15$_{-0.03}^{+0.18}$ & $-0.31_{-0.07}^{+0.11}$ & 0.81$_{-0.01}^{+0.03}$ \\
2426\_1736 & 7.822 & $-$20.65$\pm$0.14 & $-2.16 \pm 0.15$ & 2.81$_{-0.67}^{+0.64}$ & 14.16$_{-0.44}^{+0.32}$ & 0.22$_{-0.09}^{+0.19}$ & $-0.36_{-0.19}^{+0.23}$ & 0.81$_{-0.06}^{+0.06}$\\
1747\_1081 & 7.838 & $-$21.59$\pm$0.14 & $-1.63 \pm 0.06$ & 1.82$_{-0.64}^{+0.36}$ & 4.37$_{-0.35}^{+0.25}$  & --- & --- & 0.50$_{-0.08}^{+0.18}$\\
1747\_902 & 7.905 &  $-$21.10$\pm$0.19 & $-2.15 \pm 0.09$ & 2.00$_{-0.48}^{+0.30}$ & 16.92$_{-0.38}^{+0.31}$ & 0.16$_{-0.05}^{+0.10}$ & $-0.32_{-0.12}^{+0.23}$ & 0.81$_{-0.04}^{+0.04}$\\
2426\_1655 & 8.030 &  $-$20.68$\pm$0.19 & $-2.05 \pm 0.13$ & 3.44$_{-0.87}^{+0.72}$ & 27.67$_{-0.65}^{+0.59}$ & 0.24$_{-0.09}^{+0.20}$ & $-0.32_{-0.16}^{+0.43}$ & 0.63$_{-0.04}^{+0.04}$\\
2426\_169-n & 8.205 &  $-$21.45$\pm$0.09 & $-2.50 \pm 0.17$ & 1.42$_{-0.33}^{+1.14}$ & 13.13$_{-0.48}^{+0.60}$ & --- & --- & 0.73$_{-0.13}^{+0.37}$\\
1747\_732 & 8.226 &  $-$21.47$\pm$0.15 & $-2.21 \pm 0.06$ & 0.97$_{-0.26}^{+0.18}$ & 10.86$_{-0.34}^{+0.35}$ &  --- & $-0.17_{-0.13}^{+0.57}$ & 0.88$_{-0.10}^{+0.10}$\\
2426\_169 & 8.230 &  $-$22.34$\pm$0.06 & $-2.02 \pm 0.10$ & 2.10$_{-0.26}^{+0.36}$ & 66.07$_{-0.73}^{+0.59}$ & 0.19$_{-0.05}^{+0.04}$ & 0.43$_{-0.18}^{+0.13}$ & 0.73$_{-0.02}^{+0.03}$\\
1747\_199 & 8.316 &  $-$21.28$\pm$0.17 & $-2.21 \pm 0.06$ & 1.47$_{-0.16}^{+0.26}$ & 11.05$_{-0.25}^{+0.33}$ & 0.13$_{-0.12}^{+0.04}$ & $-0.26_{-0.09}^{+0.08}$ & 0.67$_{-0.05}^{+0.07}$\\
2426\_1777 & 8.440 &  $-$20.44$\pm$0.41 & $-2.37 \pm 0.36$ & 0.89$_{-0.40}^{+0.58}$ & 12.85$_{-0.39}^{+0.43}$ & --- & --- & 0.71$_{-0.10}^{+0.09}$\\
2426\_1130 & 8.490 &  $-$20.69$\pm$0.22 & $-2.37 \pm 0.36$ &  0.66$_{-0.66}^{+0.57}$ & 5.56$_{-0.55}^{+0.51}$  & --- & --- & 0.71$_{-0.14}^{+0.40}$\\
\enddata
\tablecomments{Column 1: The galaxy named by observing program followed by ID. Columns 2--4 show the redshift, M$_\mathrm{UV}$ and $\beta_{\rm{UV}}$ values from \citetalias{roberts-borsani_borg-jwst_2025}. Columns 5--6 show the \oiiS\,$\lambda \lambda$3726,29 and \oiii\,$\lambda$5007 emission line fluxes and errors in cgs units as calculated in \S \ref{emission-fit}. Column 7 has the H$_\delta$/H$_\beta$ Balmer ratios and errors. Columns 8--9 have the calculated \neiii\,$\lambda$3870/\oiiS\,$\lambda \lambda$3726,3729 (Ne3O2) and  \oiii\,$\lambda$5007/H$\beta$ (R3) line ratios and errors.}
\label{tabla}
\end{deluxetable*}

\begin{figure*}[t!]
\centering
\includegraphics[width=0.9\linewidth]{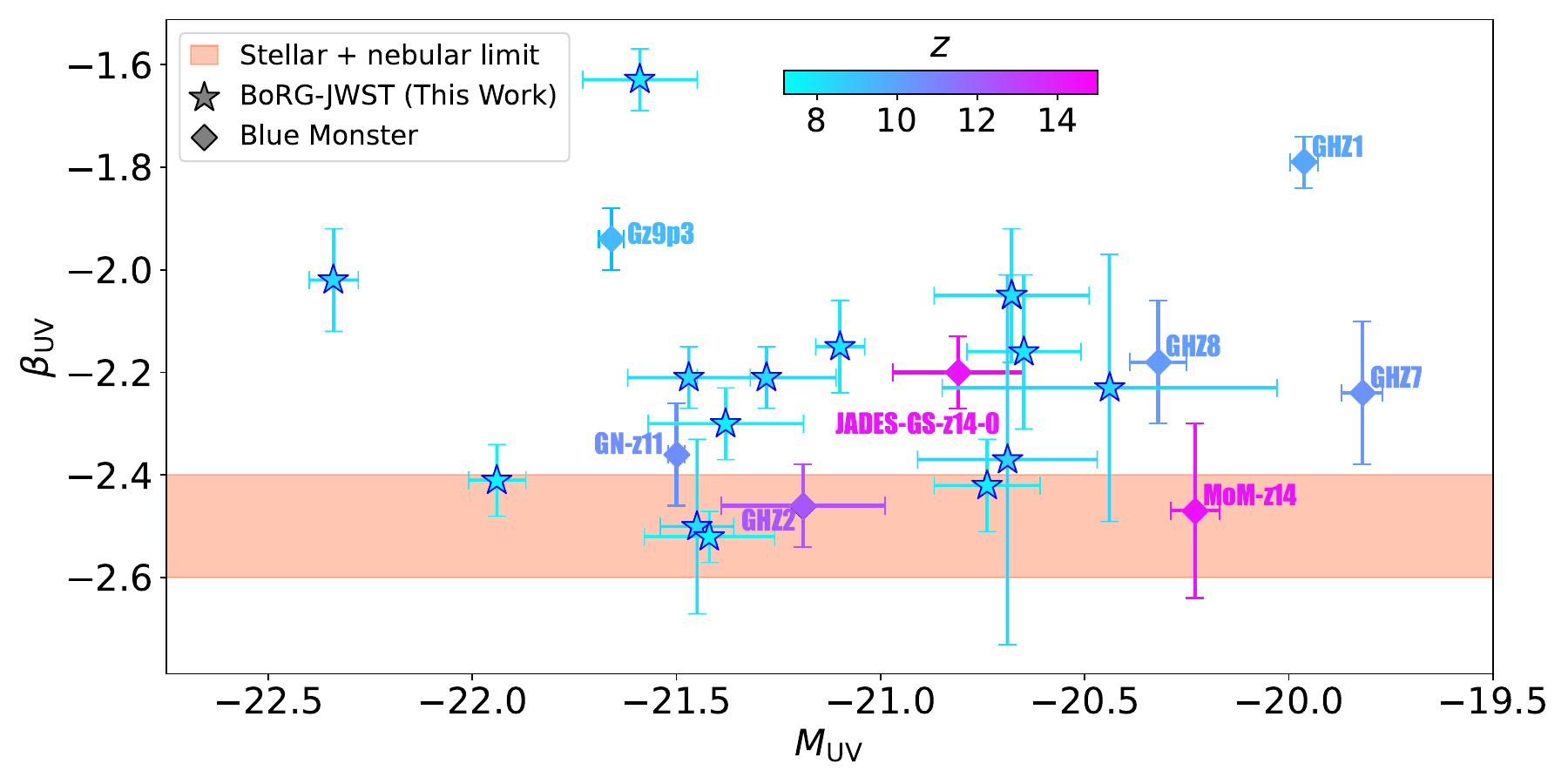}
\includegraphics[width=0.9\linewidth]{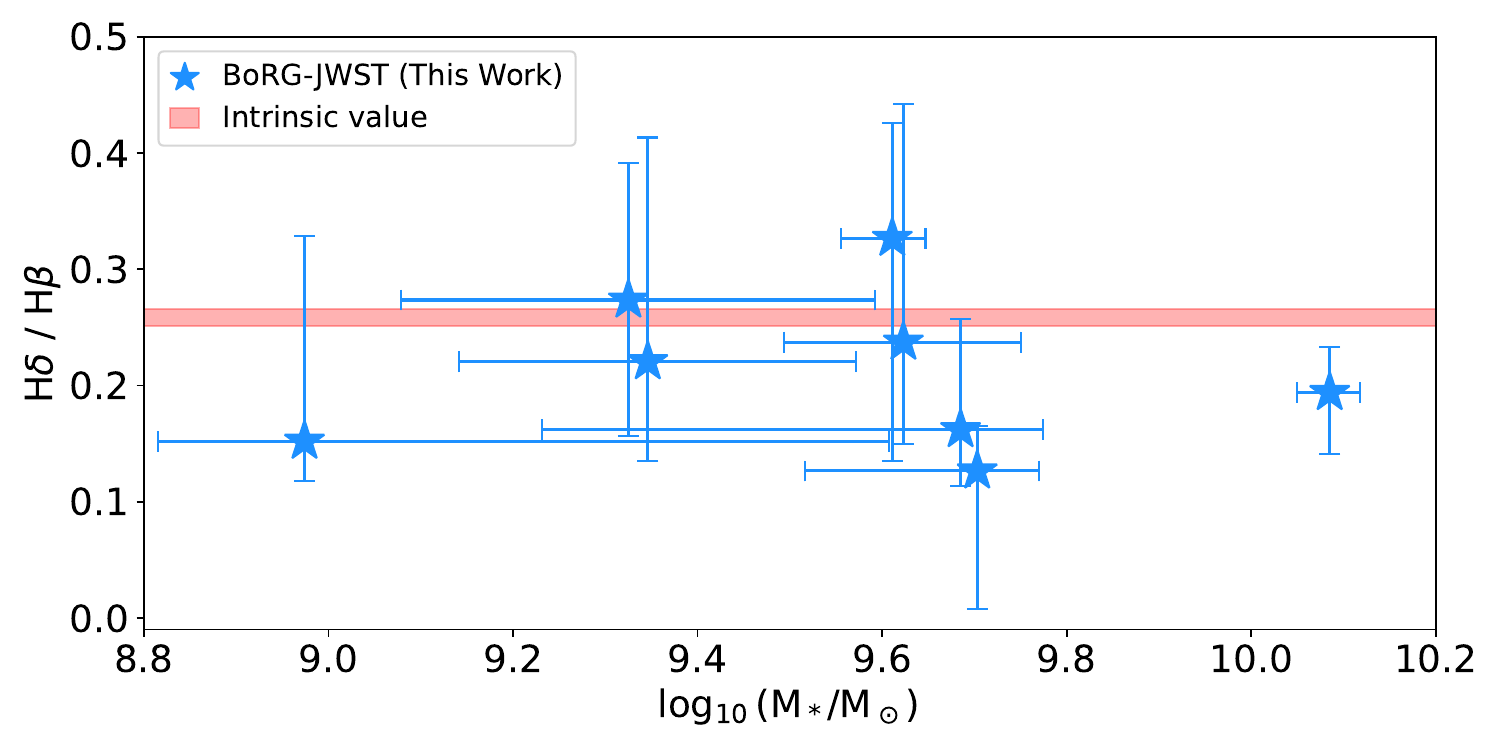}
\caption{Evaluation of the dust content. \textit{Top}: UV continuum $\beta$ slope vs. UV magnitude of BoRG-\jwst\ galaxies (stars) and spectroscopically confirmed blue monsters (diamonds). The sources are color-coded by redshift `$z$'. The orange shaded region shows the minimum value expected for $\beta$ assuming a standard stellar population with no dust and maximum nebular contribution to the emission as described in \citealt{cullen_ultraviolet_2024}. \textit{Bottom:} The Balmer decrement H$\delta$/H$\beta$ of the BoRG-\jwst\ galaxies (stars) shows that the line ratios are in agreement within 1-$\sigma$ to the intrinsic value assuming case B recombination in \citet{osterbrock_astrophysics_2006}.}
\label{fig:no-dust}
\end{figure*}

\subsection{Spectral Fitting}\label{sed-fit}
We perform spectral fitting on the full prism spectrum for each of the BoRG-\jwst\ galaxies using the Grism SED fitter code \texttt{gsf} developed by members of our team \citep[ver1.9.2][]{morishita_massive_2019}. This code is flexible to model multiple bursts or sudden declines in star formation history (SFH) using top-hat bins in age intervals. We can thus explore non-parametric SFHs to see evidence for bursty episodes of star formation that have been proposed as responsible mechanisms for the strong UV-brightness (M$_{\rm UV}$) and high stellar masses of BoRG-\jwst\ galaxies \citep{rojas-ruiz_borg-jwst_2025}, and of blue monsters \citep{ferrara_stunning_2023,shen_impact_2023}. We also tested exponentially declining SFHs following the \texttt{gsf} $\tau-$model procedure; however, these models fail to reproduce the observed spectra appropriately. The resulting fits are consistently poorer than the non-parametric solutions, with Bayesian Information Criterion differences of $\Delta\mathrm{BIC}\approx 400-7000$ in favor of the non-parametric model. We therefore continue the analysis using the non-parametric SFH results.

The spectra are modeled by finding the best combination of composite stellar populations that match the observations. The model templates are generated from the \texttt{fsps} library \citep{conroy_propagation_2009,foreman-mackey_emcee_2013} taking metallicity and dust attenuation as free parameters with $\log Z_*$/$Z_{\odot} \in $ [$-2,0$] in increments of 0.1, and $A_V \in $ [$0,4$]. We make a list of redshifts to sample the star formation history starting more densely closer to the age of the universe at the time the galaxy is observed and more sparsely earlier in time, in order to capture recent episodes with appropriate time resolution. In practice, this is achieved starting from the age of the universe at the time of observation and halving the interval multiple times. For example, a galaxy at $z=8.03$ with a corresponding age of the universe of $624$ Myr, has a grid of ages [19, 39, 78, 156, 312, 624] Myr. We also add a nebular component characterized by $\log U \in [-3, -0.5]$ in increments of 0.5, which is included in the fitting process with a free amplitude parameter. The metallicity of the nebular component is set to that determined for the stellar template at the youngest age bin.

Through the spectral fitting we obtain the SFHs and the stellar mass of the galaxy ($\log M_*$) at the observed redshift, and at different lookback times. We also post-process the posterior SFH and recalculate the M$_{\rm UV}$ that the galaxy would have had if observed at earlier redshifts. This enables a more direct comparison to the observed M$_{\rm UV}$ of the $z\gtrsim10$ UV-bright galaxies in the literature. For this calculation, we linearly sum the SED template at each lookback time to account for older stellar populations. At each lookback time, we scale the template based on the amount of stars that form in the previous age. This results in a cumulative SED that is used to calculate the UV luminosity and then the M$_{\rm UV}$ at each lookback time (see \S\ref{bursts}). We adopt the 16th and 84th percentiles as lower and upper bounds on the uncertainty, respectively, of all these best-fit parameters.

\section{Results and Discussion}
Below we present results on the analysis of the BoRG-\jwst\ galaxies that confirm their analog nature to $z\gtrsim10$ blue monsters, as their distribution of UV luminosities is indistinguishable spanning $-22.5 < {\rm M}_{\rm UV} < -19.5$. For comparison, we include the galaxies Gz9p3 at $z=9.3127$ \citep{castellano_early_2023,boyett_massive_2024}; GN-z9p4 at $z=9.380$\citep{schaerer_discovery_2024}; GHZ1 at $z=9.875$, GHZ8 at $z=10.231$ and GHZ7 at $z=10.43$ \citep{napolitano_seven_2025}; GN-z11 at $z=10.603$ \citep{oesch_remarkably_2016,bunker_jades_2023}; GHZ2 at $z=12.34$ \citep{castellano_early_2022,castellano_jwst_2024}; JADES-GS-z14-0 at $z=14.18$ \citep{carniani_spectroscopic_2024,schouws_detection_2025}, and MoM-z14 at $z=14.44$ \citep{naidu_cosmic_2025}.

\subsection{Lack of dust extinction}\label{lack-dust}

The $\beta_{\rm UV}$ continuum slope is a sensitive probe of dust attenuation and stellar populations when direct measurements of dust content are not available. As an example, \citet{cullen_ultraviolet_2024} performed an analysis of the $\beta_{\rm UV}$ slope in photometric samples of $z\gtrsim8$ galaxies using an extended wavelength baseline afforded by the combination of \jwst/NIRCam and COSMOS/UltraVISTA. Their analysis on the evolution of $\beta_{\rm UV}$ with redshift showed that $z>10.5$ galaxies had on average values of $\beta_{\rm UV}-2.4$, in agreement with values attributable to standard stellar populations with minimum dust content and maximum contribution of the nebular continuum to the emission of these galaxies \citep[e.g.,][]{cullen_first_2017,reddy_hduv_2018}. However, several galaxies at higher redshifts display slopes bluer than the $\beta_{\rm UV}\sim-2.6$
limit for star-forming galaxies with negligible dust attenuation \citep{chisholm_far-ultraviolet_2022}, thus questioning the mechanisms producing the UV continuum emission in blue monsters. 

Here we have the opportunity to spectroscopically verify measurements of $\beta_{\rm UV}$ in the most luminous galaxies at $z\sim7-9$, reducing potential biases in the measurements from photometric scatter and possible emission line contributions. We take the $\beta_{\rm UV}$ continuum slope values measured by \citetalias{roberts-borsani_borg-jwst_2025} and plot them as blue stars in the top panel of Figure \ref{fig:no-dust}, also adding the measurements of the blue monsters color-coded by redshift. We find that both populations of intrinsically UV-luminous galaxies tend to have little dust and be within $2-\sigma$ of the value for zero dust attenuation, with an average $\beta_{\rm UV}=-2.23\pm0.23$ for the BoRG-\jwst\ galaxies, and $\beta_{\rm UV}=-2.20\pm0.24$ for the spectroscopically confirmed $z\gtrsim10$ galaxies,  with the lowest cases consistent with the extreme $\beta_{\rm UV}\lesssim-2.6$ value.

The spectroscopic BoRG-\jwst\ data set affords a unique and practical opportunity to verify the ``no-extinction" hypothesis proposed to explain the extreme UV luminosities of the $z>10$ galaxy population. The ``Balmer decrement" has proven to be an efficient method to infer
the dust content of galaxies \citep[e.g.,][]{glazebrook_measurement_1999,reddy_multiwavelength_2008,madau_cosmic_2014,shapley_rest-frame_2003,sandles_jades_2024}.
The intrinsic ratios of relevant Balmer lines are calculated under ideal conditions of a dust-free gas cloud, assuming Case B recombination and a typical temperature of $\sim10,000$ K \citep{osterbrock_astrophysics_1989,osterbrock_astrophysics_2006}. Although the Balmer decrement is traditionally estimated from the brightest Hydrogen Balmer lines H$\alpha$/H$\beta$, the H$\alpha$ line is redshifted out of the NIRSpec observations, while the prism's low spectral resolution (R$\sim$100) prevents us from resolving $H\gamma$ emission from the auroral \oiii$ \lambda4363$ emission line. Thus, we rely on the H$\delta$/H$\beta$ ratio for our Balmer decrement analysis. 

We use the nebular emission line analysis code PyNeb \citep{luridiana_pyneb_2015} to calculate the intrinsic values of the H$\delta$/H$\beta$ ratio following the Case B recombination. The code uses the line emissivity intensities from \citet{storey_recombination_1995} and we adapt electron densities in the range $10^{1-6}$ cm$^{-3}$ and temperatures $5,000 -20,000$~K. This yields intrinsic values for H$_\delta$/H$_\beta = 0.251- 0.266$. The measured Balmer ratios for the BoRG-\jwst\ galaxies with detected H$_\delta$ and H$_\beta$ emission at S/N $>2.5$ are presented in Table~\ref{tabla}. We show these ratios as a function of stellar mass $\log$M$_*$/M$_\odot$ in the bottom panel of Figure~\ref{fig:no-dust}, with the intrinsic ratios highlighted in red. We stack the spectra of the BoRG-\jwst\ galaxies with significant  H$_\delta$ and H$_\beta$ detections an measured an average ratio H$_\delta$/H$_\beta =0.216^{+0.056}_{-0.040}$, consistent within 1-$\sigma$ of the theoretical ratios calculated. Assuming Calzetti's law, The corresponding extinction is $A_V= 1.05^{+1.13}_{-1.05}$ mag, suggesting that while some dust cannot be excluded, the data are consistent with near-zero extinction. There is no evidence of correlation with galaxy mass. The Balmer decrement further supports the hypothesis of negligible dust extinction in the BoRG-\jwst\ sample. While such confirmations do not preclude dust production at such early times, the virtually dust-free interstellar medium (ISM) conditions presented here suggest that some mechanism is required to either destroy the dust grains or have them ejected from the ISM for them to remain undetected.

\subsection{Nature of the sources: AGN- vs. SF-powered}\label{agn-sf}
 
One of the suggestions to explain such efficient dust depletion in hyper-luminous galaxies is through ejection by outflows from either strong UV radiation pressure of young, massive metal-poor stars, or from an AGN \citep{ziparo_blue_2023}.

Previous work on the selection of BoRG sources utilized morphological analyses to either limit or exclude obvious AGN. \citet{morishita_superborg_2020} characterized the expected number density of quasar-like point sources from pure-parallel observations, suggesting $>100\times$ survey volumes were required to identify quasar-like objects of comparable luminosities to the UV-selected galaxies. Other studies instead applied more direct efforts such as stellarity cuts to exclude the most obvious AGN \citep{trenti_brightest_2011,calvi_bright_2016,rojas-ruiz_probing_2020,bagley_bright_2024}, potentially at the cost of sample completeness.

Once again, despite the challenging endeavour of identifying AGNs, the BoRG-\jwst\ spectra provide us with direct measurements to test for obvious spectral signatures of AGN and the theory of dust-ejection by AGN-driven winds. Traditional ``BPT'' diagrams rely on the detection of strong H$\alpha$\,$\lambda$6563 emission \citep[e.g.,][]{baldwin_classification_1981,kewley_theoretical_2001,kauffmann_host_2003,feltre_nuclear_2016}, which is beyond the wavelength coverage of NIRSpec at $z>8$. Rest-frame UV diagnostics from high-ionization such as \civ\,$\lambda \lambda$1548,1550, \ciii$\lambda$1909, and \heii$\lambda$1640 lines provide a useful alternative, however these are mostly undetected in the relatively shallow individual spectra of BoRG-\jwst\, and indeed in most individual high-redshift NIRSpec spectra from the literature.

\begin{figure}[t!]
\centering
\includegraphics[width=\linewidth]{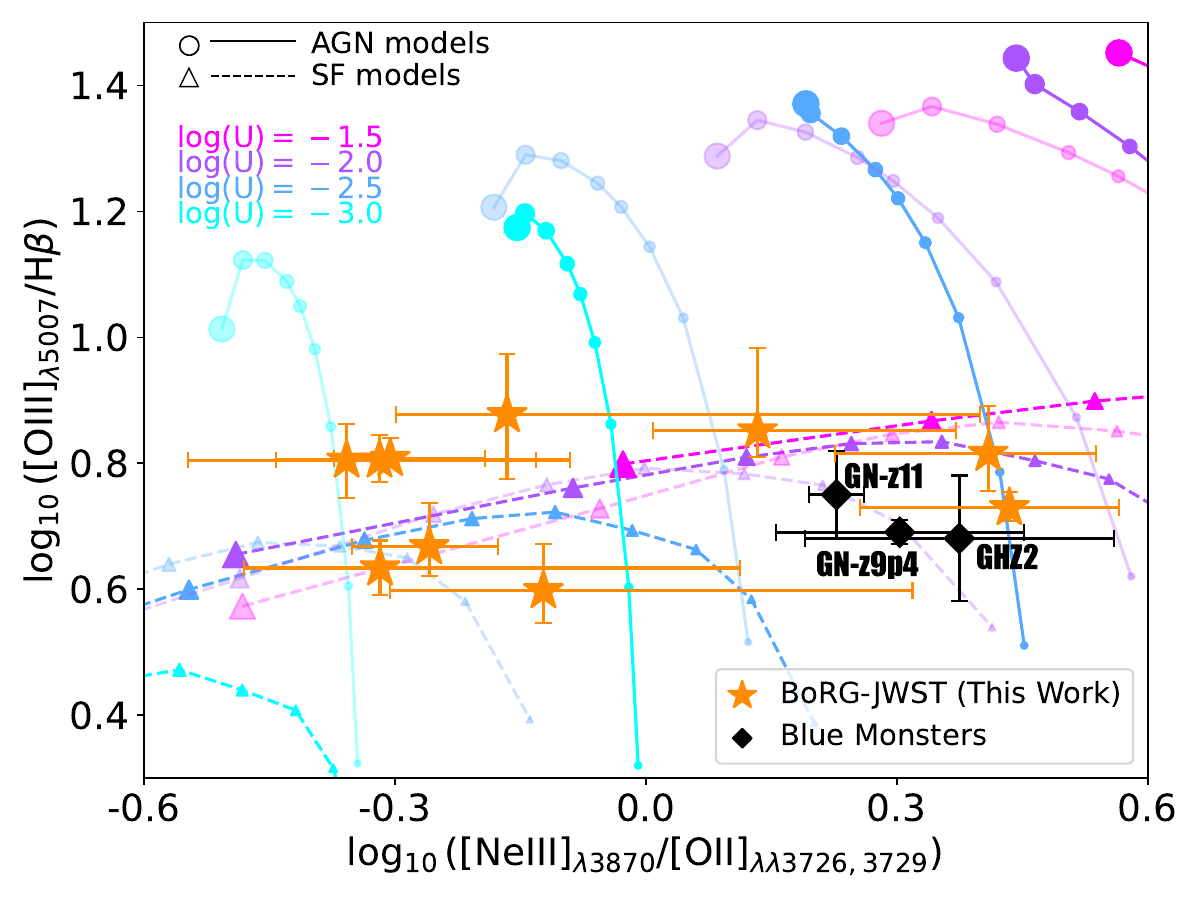} \includegraphics[width=\linewidth]{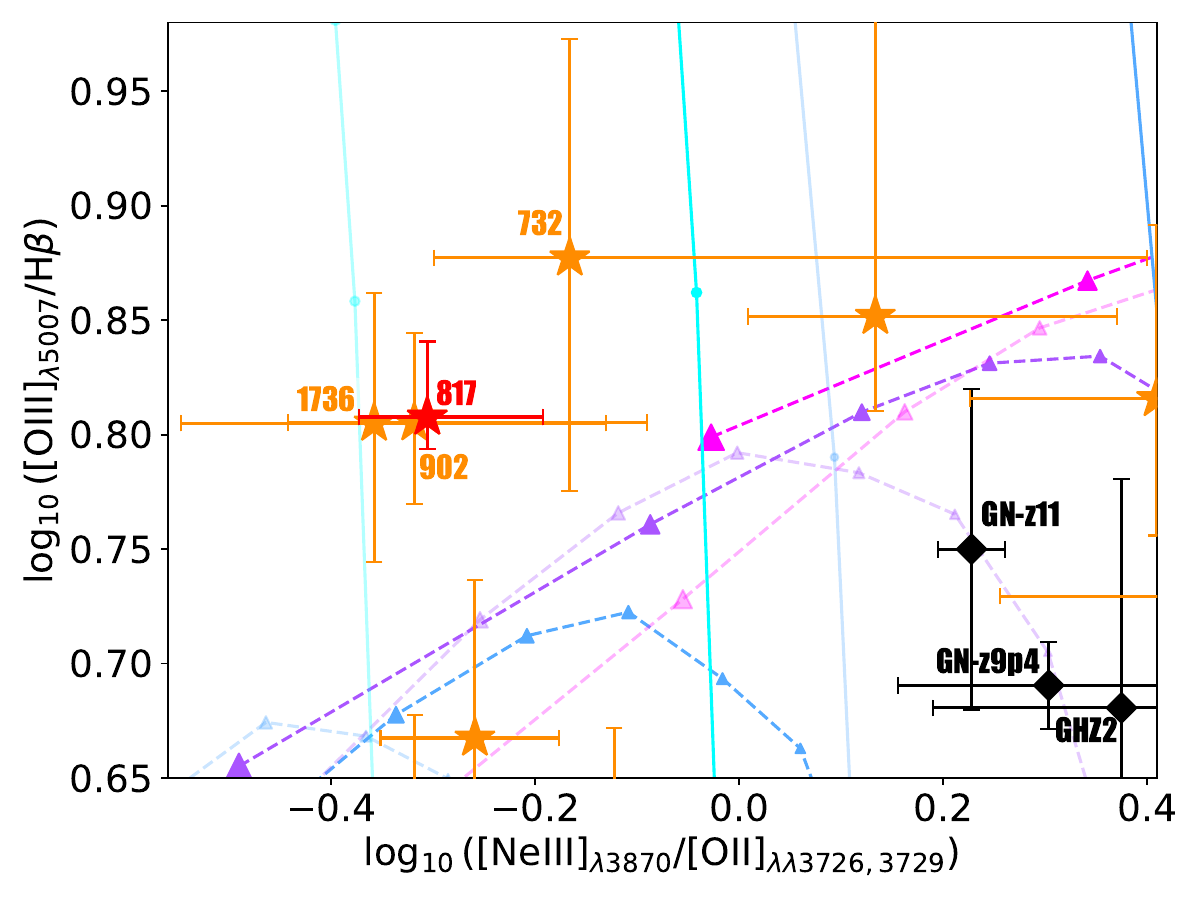}
\caption{OHNO diagnostic for AGN (contiguous lines with circles) and star-forming galaxies (dashed-lines with triangles). \textit{Top}: 
The models are color-coded by ionization parameter $\log(U)$ = $[-1.5, -3.0]$ with different color intensities for electron densities $n_e= 10^4$ cm$^{-3}$ (bright) $10^3$ cm$^{-3}$ (fainter). The symbols increase in size from subsolar to solar metallicities ($0.05<Z_{\mathrm{gas}}/Z_\odot \leq1.0$). The measurements for the BoRG-\jwst\ galaxies are shown as orange stars and three UV-luminous galaxies at $z>10$ as black diamonds. \textit{Bottom:} Zoom in of the previous Top figure in the area where AGN solution is more exclusive, although within 1$\sigma$ (1747\_732) and 2$\sigma$ (1747\_902, 2426\_1736) of the star forming model solution. Source 1747\_817 highlighted in red is marginally consistent with AGN classification, lying 3$\sigma$  and 7.5$\sigma$ from star-forming models in the Ne3O2 and R3 directions, respectively.}
\label{fig:agn-sf}
\end{figure}

Nonetheless, a number of useful rest-optical alternatives exist, making use of moderately strong emission lines to differentiate between star-formation and AGN contributions. In particular, we follow the diagnostics showcased by \citet{calabro_evidence_2024} who utilize well-known ratios of \neiii\,$\lambda$3870/\oiiS\,$\lambda \lambda$3726,3729 (Ne3O2) vs. \oiii\,$\lambda$5007/H$\beta$ (R3), the ``OHNO'' diagnostic. We present the measured line ratios and uncertainties of our sample in Table~\ref{tabla}; galaxies 1747\_1081,  2426\_1130, 2426\_169-n, and 2426\_1777 do not have \neiii\,$\lambda$3870 emission detected and are therefore not presented. The OHNO diagnostic plot is shown in Figure~\ref{fig:agn-sf}, where we also plot the AGN models (contiguous lines with circles) and star-forming models (dashed lines with triangles) from \citet{calabro_near-infrared_2023}, for comparison. The models are color-coded by ionization parameter (ranging from $\log U = -3.0$ to $-$1.5), with two electron densities for each $\log U$, $n_e = 10^4$ cm$^{-3}$ (brighter color) and $10^3$ (fainter color). The metallicity is represented by the symbols that increase in size from 0.05 to solar metallicities. We add for reference a few blue monsters for which the relevant emission lines or line ratios have already been measured and reported; GN-z9p4 \citep{schaerer_discovery_2024}, GN-z11 \citep{bunker_jades_2023,alvarez-marquez_insight_2025}, and GHZ2 \citep{calabro_evidence_2024,zavala_luminous_2025}.

We find that the BoRG-\jwst\ measurements span either the low end of AGN ionization models with $\log U < -2.0$, or virtually the full range of ionization parameters for star-forming models. We find that a subset of these (four sources) possibly prefer the AGN models with low ionization $\log U = -3.0$, as highlighted in Figure \ref{fig:agn-sf}, bottom.  However, the ratios are within 1$\sigma$ (source 1747\_732), or 2$\sigma$ (1747\_902, and 2426\_1736) of the star-forming models. Source 1747\_817 can marginally be classified as an AGN with this diagnostic given that its Ne3O2 ratio deviates by 3$\sigma$ from the star-forming model with $\log U = -1.5$, $n_e = 10^4$ cm$^{-3}$, and solar metallicity, while the R3 ratio differs by 7.5$\sigma$ from the model at $\log U = -2.0$, similar density, and near-solar metallicity $Z_{\mathrm{gas}}/Z_\odot \sim 0.7-1.0$. A low-ionization AGN is a more plausible explanation than a solar-metallicity star-forming galaxy at $z=7.556$, although at least one such high-metallicity system has been reported in the literature (see \cite{shapley_aurora_2025}).

However, we see no evidence for additional or unambiguous signatures suggesting strong or dominant AGN contributions. All these sources, except for 1747\_732, present detection of strong \ciii\,$\lambda \lambda$1907, 1909 at S/N $>3.0$, with 1747\_817 having emission of the blended \oiii\,$\lambda 1660$,$ \lambda 1666$ at S/N$ = 5.1$, as is typical for high-redshift star-forming sources \citep{roberts-borsani_between_2024}. These emission lines typically trace strong ionization from either low-metallicity stellar population or AGN activity, with their equivalent widths serving as a key diagnostic in conjunction with other, higher-ionization UV lines \citep[e.g.,][]{stark_ly_2017,nakajima_vimos_2018,fevre_vimos_2019}. However, our spectral resolution is insufficient to reliably deblend the components or to robustly estimate the equivalent width.

Thus, while our sample are consistent with the range of star-forming models with low metallicity and strong ionization parameters, we cannot rule out moderate AGN contributions with the present data set. Deeper and higher spectral resolution data will be required for definitive conclusions.

\begin{figure*}[ht!]
\centering
\includegraphics[width=\linewidth]{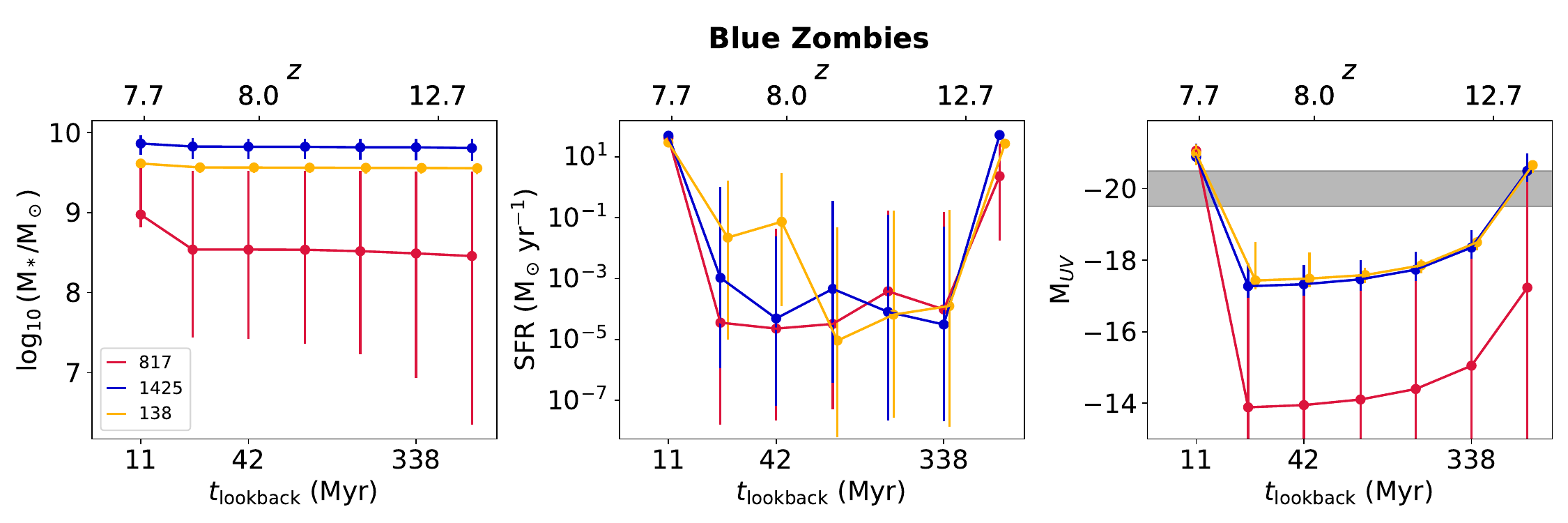}
\includegraphics[width=\linewidth]{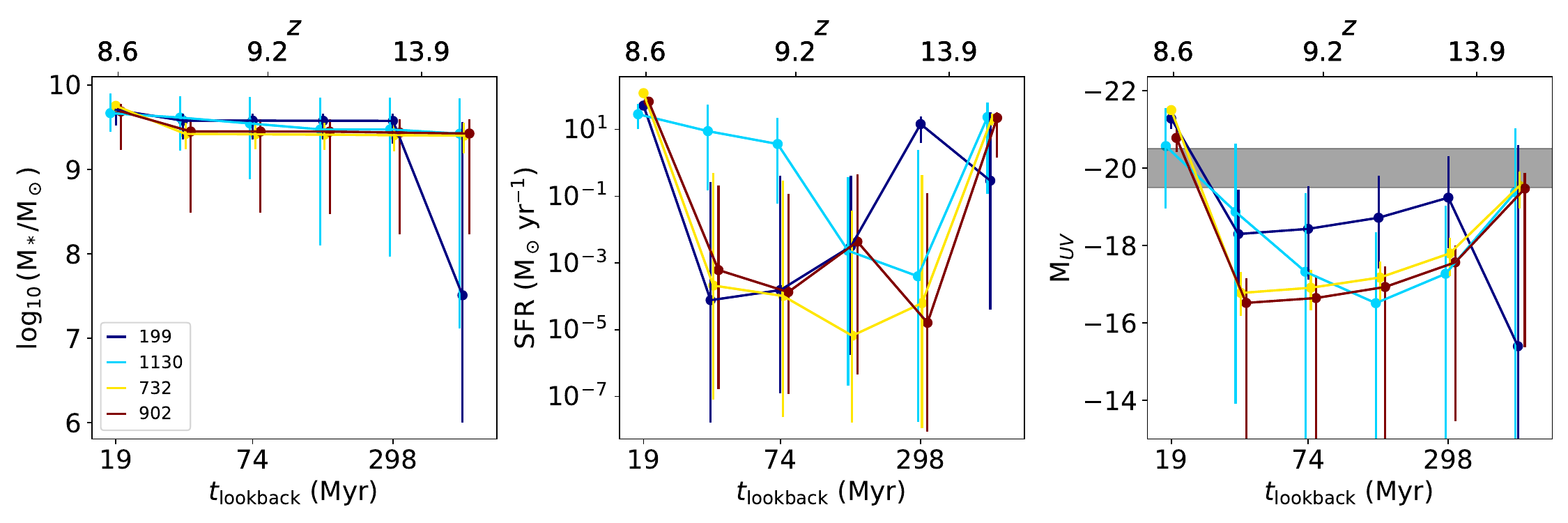}

\caption{Inferred evolution of the BoRG-\jwst\ galaxies using spectral fitting. \textit{Left}: Stellar mass growth as a function of lookback time from the epoch of observation and corresponding redshift. \textit{Center:} Star formation rate, displaying intense bursts of star formation. \textit{Right:} Cumulative M$_{\rm{UV}}$ luminosity as a function of lookback time/redshift; the gray shaded region shows the limit at which the galaxy approaches M$_{\rm{UV}}\sim -20.$ The galaxies in this first category of ``Blue Zombies'' are characterized by having a first burst at $z\gtrsim 12$ reaching M$_{\rm{UV}}\lesssim -20$, lull over time, and then have a recent burst closer to observation ($z\simeq7-9$).
}
\label{fig:bursts}
\end{figure*}

\subsection{Bursty Episodes of Star Formation}\label{bursts}

Stochastic (or ``bursty'') star formation histories have been proposed as a key feature to explain the extreme UV luminosities of the blue monsters. These bursty episodes of star formation could arise as a consequence of multiple feedback processes typical in early galaxies ($z>7$) driven by supernovae \citep[e.g.,][]{el-badry_breathing_2016,mirocha_balancing_2023,shen_impact_2023,sun_bursty_2023} or outflows from stellar winds \citep[e.g.,][]{gelli_impact_2024,carnall_first_2023}.  
Based on the model explored by \citet{ferrara_eventful_2024} for JADES-GS-z14-0 at $z=14.18$, blue monsters have been suggested to have undergone a short and intense burst of star formation in the last $\sim40$~Myr before observation, accompanied by outflows revealing UV-bright and young stellar populations. Before this, other obscured or smoldering bursts of star formation can last 30-40~Myr and are separated by quenching phases of $\sim 20$~Myrs. Thus, from the birth of the first stars, star formation and thus the stellar mass increases with time in patches rather than smoothly. These bursts in the evolution should be identifiable in the SFH, however constraints from the rest-frame optical --where longer-lived stars emit the majority of their light-- are fundamental for accurate constraints (e.g., \citealt{witten_rising_2025}) over more extended timescales. At $z\gtrsim10$, such constraints are beyond the wavelength range of NIRSpec and NIRCam, and impractical with MIRI, highlighting the usefulness of the hyper-luminous BoRG-\jwst\ objects at $z\simeq7-9$.

\begin{figure*}[ht!]
\centering

\includegraphics[width=\linewidth]{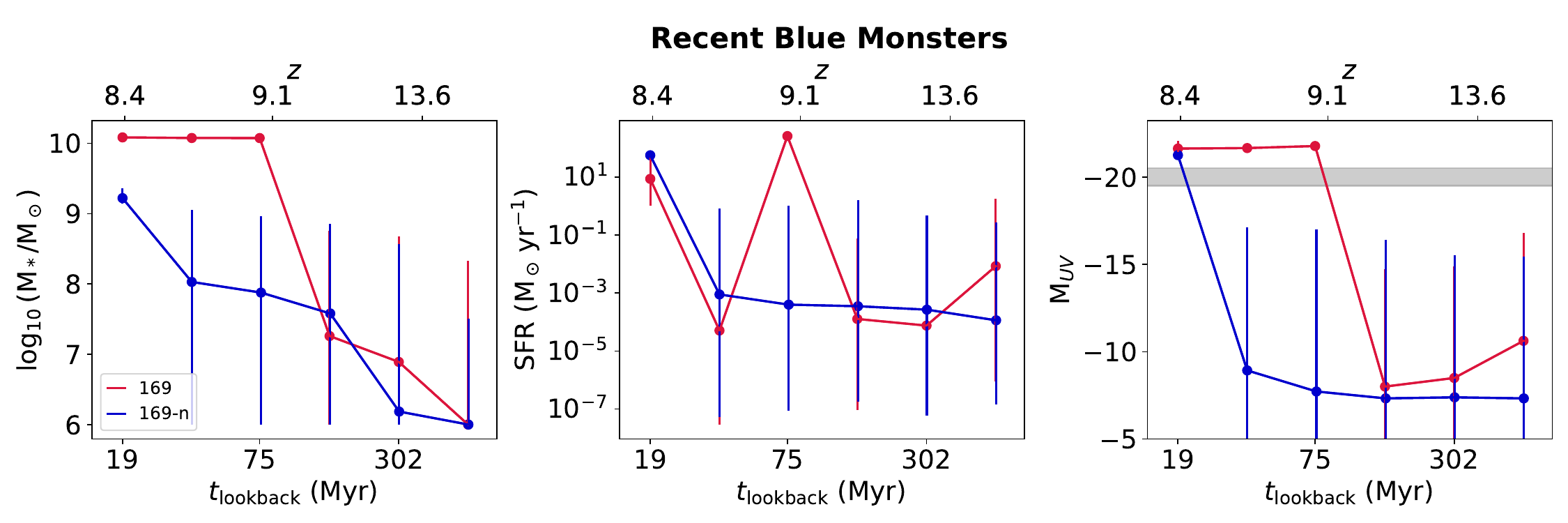}
\includegraphics[width=\linewidth]{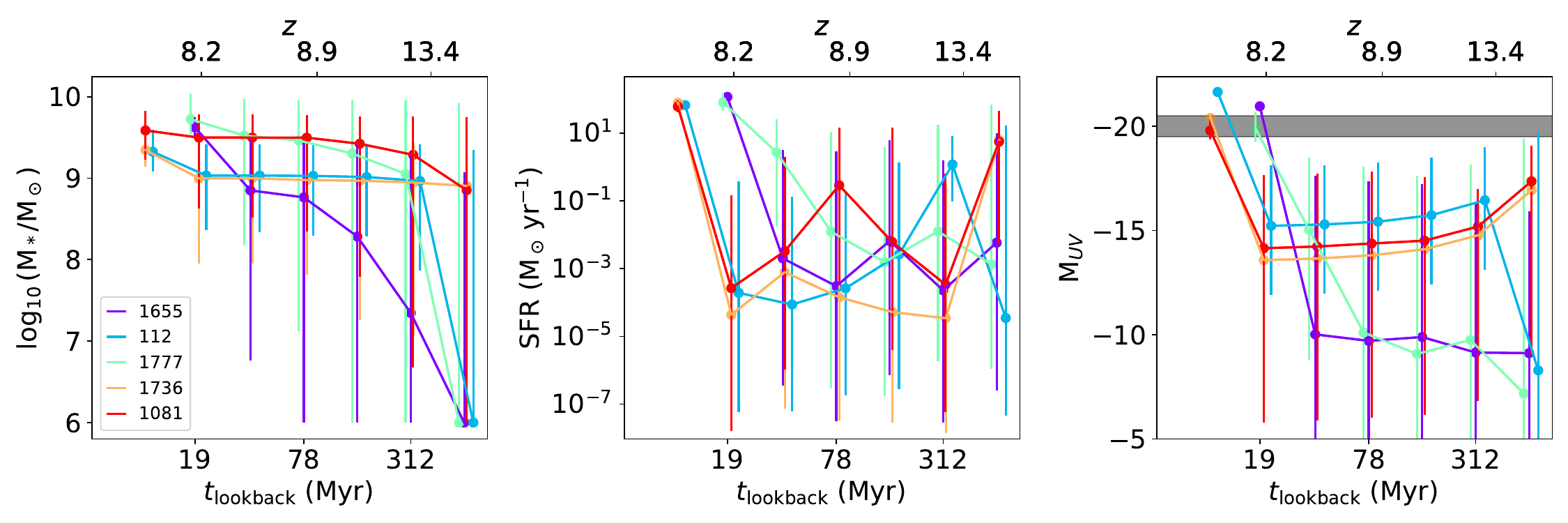}

\caption{Same as Figure \ref{fig:bursts} for the other half of the sample denominated ``Recent Blue Monsters'' characterized by a recent burst in star formation reaching a brightness M$_{\rm{UV}}\lesssim -20$ closer to observation.}
\label{fig:bursts2}
\end{figure*}

With rest-UV-to-optical spectra in hand, we model the SFH of the BoRG-\jwst\ sample using the \texttt{gsf} spectral fitting code, to identify the presence (or otherwise) of bursty episodes of star formation. In Figures~\ref{fig:bursts} and \ref{fig:bursts2} we present the stellar mass, star formation rate, and cumulative M$_{\rm{UV}}$ as a function of lookback time for each of the BoRG-\jwst\ galaxies. We group the results into two categories according to the redshift at which the first burst surpasses the M$_{\rm{UV}}\sim-20$ brightness. In the first category --named ``Blue Zombies" because they become blue monsters a first time, they fade, and then become blue monsters again (Figure~\ref{fig:bursts})-- we find that half the galaxy sample undergoes a burst within the first 300~Myr of formation, corresponding to $z\gtrsim12$: 1747\_817, 1747\_1425, 1747\_138, 1747\_199, 2426\_1130, 1747\_732, and 1747\_902 . This early burst builds up substantial stellar masses of $\log_{10}$M$_*$/\Msun$\gtrsim8$, as observed by the mass-weighted stellar ages in Table~\ref{tablemass}. The mass continues to grow smoothly until a second burst occurs within the final 20~Myr before observation, which agrees with the evolutionary model proposed by \citet{ferrara_eventful_2024}.

In the second category --named ``Recent Blue Monsters"-- is the rest of the sample that is dominated by recent bursts within 20~Myrs of the time of observation (see Figure~\ref{fig:bursts2}). A special case is galaxy 2426\_169 which experiences a very strong ($260$ \Msun yr$^{-1}$) first burst after $\sim530$~Myr ($z\sim9.1$). After the burst, the star formation rate declines but the galaxy remains bright M$_{\rm{UV}}<-20$, aided by a secondary burst closer to the time of observations. This case shows how the effects of multiple bursts can contribute to the observed UV luminosity, as proposed by e.g., \citet{mason_brightest_2023,narayanan_outshining_2024}. Unsurprisingly, this galaxy reaches the largest stellar mass in the sample at $\log_{10}$M$_*$/\Msun$ = 10.09_{-0.04}^{+0.03}$, and brightest luminosity M$_{\rm{UV}}=-22.34\pm0.06$. This galaxy was proposed as a neighbor in the same environment of galaxy 2426\_169-n \citep{rojas-ruiz_borg-jwst_2025}. Evaluating the SFH of both galaxies, we find they are remarkably different with 2426\_169-n having a smooth evolution and no significant changes at the same lookback time as the strong starburst for 2426\_169 (see Figure \ref{fig:bursts2}, Top panel). 

\begin{deluxetable}{lccc}[h]
\tablecaption{Star Formation History for Galaxy Categories}
\tablewidth{700pt}
\tablehead{
\colhead{ID}
& \colhead{z} & \colhead{$\log_{10}$M$_*$/\Msun} & 
\colhead{MassW Age (Myr)}
}
\startdata
\multicolumn{4}{c}{Blue Zombies} \\
\hline
1747\_817 & 7.556 & 8.97$_{-0.16}^{+0.63}$ & 216.3$_{-192.5}^{+306.1}$ \\
1747\_1425 & 7.553 & 9.86$_{-0.14}^{+0.11}$ & 595.7$_{-58.6}^{+63.5}$ \\
1747\_138 & 7.179 & 9.61$_{-0.06}^{+0.04}$ & 632.4$_{-66.2}^{+65.8}$\\
1747\_199 & 8.316 & 9.70$_{-0.19}^{+0.07}$ & 262.4$_{-99.1}^{+153.5}$\\
2426\_1130 & 8.490 & 9.66 $_{-0.22}^{+0.24}$ & 352.4$_{-288.9}^{+135.2}$\\
1747\_732 & 8.226 & 9.75$_{-0.07}^{+0.06}$ & 280.0$_{-73.8}^{+65.3}$\\
1747\_902 & 7.905 & 9.68$_{-0.45}^{+0.09}$ & 359.7$_{-268.8}^{+72.8}$\\
\hline
\multicolumn{4}{c}{Recent Blue Monsters} \\
\hline
2426\_169 & 8.230 & 10.09$_{-0.04}^{+0.03}$ & 76.7$_{-19.1}^{+21.21}$\\
2426\_169-n & 8.205 & 9.22$_{-0.08}^{+0.14}$ & 31.8$_{-16.0}^{+85.7}$\\
2426\_1655 & 8.030 & 9.62$_{-0.13}^{+0.13}$ & 45.5$_{-28.5}^{+165.9}$\\
2426\_112 & 7.337 & 9.32$_{-0.25}^{+0.27}$ & 173.8$_{-112.1}^{+243.1}$\\
2426\_1777 & 8.440 & 9.72$_{-0.19}^{+0.32}$ & 140.3$_{-119.3}^{+322.1}$\\
2426\_1736 & 7.822 & 9.35$_{-0.21}^{+0.23}$ & 253.3$_{-226.2}^{+223.0}$\\
1747\_1081 & 7.838 & 9.59$_{-0.36}^{+0.24}$ & 268.0$_{-232.5}^{+282.9}$\\
\enddata
\tablecomments{Column 1: The galaxy program followed by ID. Columns 2: The galaxy's redshift. Column 3-4: The final stellar mass calculation and mass-weighted stellar ages from \texttt{gsf} modeling.}
\label{tablemass}
\end{deluxetable}

Interestingly, galaxies 1747\_112, 1747\_817, 2426\_1736, 2426\_1655, and 2426\_169-n have evidence for the most recent burst to have high specific star formation rates (sSFR) exceeding 25 Gyr$^{-1}$, which is compatible with the scenario by \citealt{ferrara_super-early_2024} advocating for dust ejection via radiation pressure through a dust-driven outflow. Galaxy 1747\_817 has the highest sSFR in the sample at sSFR  $=45^{+111}_{-18}$ Gyr$^{-1}$, in agreement with galaxy JADES-GS-z14-0 \citep{ferrara_eventful_2024}. The analysis presented here, utilizing constraints from both the rest-frame UV to optical, confirm the bursty nature of the hyper-luminous galaxy population, whose stochasticity clearly serve to temporarily and periodically enhance their UV luminosities to extreme levels.

To further assess whether the Blue Zombies and Recent Blue Monsters are distinct populations, we perform an additional analysis as described below. We compared the mass-weighted stellar age distributions as they quantify when the bulk of the stellar mass was formed, allowing us to test whether Blue Zombies experienced earlier mass assembly due to an earlier burst of star formation, than the Recent Blue Monsters. We performed an Anderson-Darling k-sample test with Python \texttt{scipy.stats.anderson\_ksamp}, which is sensitive to differences in the tails of the distributions of small samples, such as this case with seven galaxies per population. The test indicates a statistically significant difference among the groups (p = 0.004), confirming that the two samples are drawn from distinct stellar mass assembly histories. Blue Zombies formed most of their mass earlier in the first burst, on average at $\sim380$ Myr before observation with minor bursts in the more recent $\sim20$ Myr contributing to the total stellar mass. In contrast, Recent Blue Monster galaxies formed all their mass more recently on average in the latest $\sim140$ Myr (See Table~\ref{tablemass}). Therefore, Blue Zombie galaxies systematically have older mass-weighted stellar ages than Recent Blue Monster galaxies, indicating earlier stellar mass assembly coincident with the earlier burst observed in their SFH.

\begin{figure*}[ht!]
\centering
\includegraphics[width=0.8\linewidth]{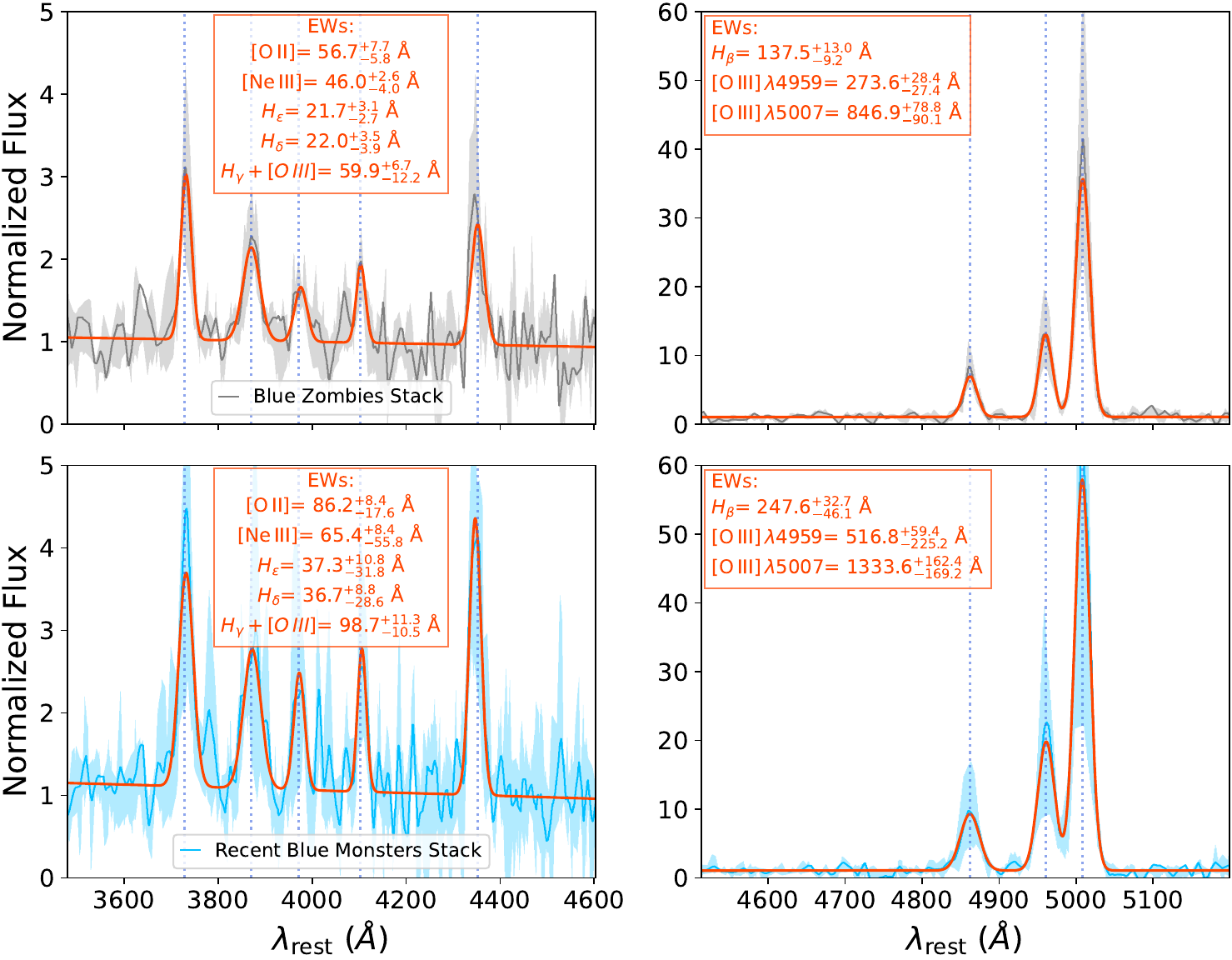}
\caption{\textit{Left:} Equivalent widths (EWs, orangered) of the \oiiS\,$\lambda \lambda$3726,3729 to H$\gamma$+\oiii\,$\lambda$4363 emission lines in the stacked spectra following the fitting procedure described in \S\ref{emission-fit}. \textit{Right:} Similar for the H$\beta$ and \oiii\,$\lambda \lambda$4959,5007 emission lines. The EWs of the Balmer lines in the Blue Zombies are consistently smaller, as expected from the presence of non-negligible underlying older stellar continuum, from the first burst of star formation, diluting the emission.}
\label{fig:z-m}
\end{figure*}

We also investigate the physical component in the spectra driving the distinct galaxy populations. The average $\beta_{\rm{UV}}$ slope for Blue Zombies is $-2.3\pm0.04$, which is consistent within $1\sigma$ with that of Recent Blue Monsters with average $\beta_{\rm{UV}}= -2.16 \pm 0.07$. Given that there is little significant differentiation based on their rest-UV spectra, we analyze the rest of the spectra more sensitive to the stellar components. For this, we stacked the spectra of both populations separately by shifting each galaxy spectrum to the rest frame, and interpolating onto a common grid of $1000-5200$~\AA. Each spectrum was normalized by the line-free continuum window 4450–4750~\AA\ before performing a mean stacking (See Figure~\ref{fig:z-m}).

We note that the Recent Blue Monster stack has higher emission line equivalent widths (EWs) of Balmer lines, consistent with more recent bursts of star formation and higher contribution to stellar mass build up, as also demonstrated with the mass-weighted stellar ages. Conversely, the lower Balmer line EWs in the Blue Zombies stack is consistent with being depressed by absorption from A stars. To show the difference between the Balmer lines of the both populations, we fitted the emission lines of the stacks following the procedure for line fitting described in \S\ref{emission-fit}. The EWs on the normalized stacks and the uncertainties were estimated via bootstrap sampling, recomputing the mean stack and EW 2000 times. In the Blue Zombie stack we find: EW$_{\rm{H}_{\epsilon}} = 21.7^{+3.1}_{-2.7}$\AA, EW$_{\rm{H}_{\delta}} = 22.0^{+3.5}_{-3.9}$\AA, EW$_{\rm{H}_{\gamma+[O\, III]}} = 59.9^{+6.7}_{-12.2}$\AA, and EW$_{\rm{H}_{\beta}} = 137.5^{+13.0}_{-9.2}$\AA. In the Recent Blue Monster stack, the EWs are systematically higher albeit larger uncertainties due: EW$_{\rm{H}_{\epsilon}} = 37.3^{+10.8}_{-31.8}$\AA, EW$_{\rm{H}_{\delta}} = 36.7^{+8.8}_{-28.6}$\AA, EW$_{\rm{H}_{\gamma+[O\, III]}} = 98.7^{+11.3}_{-10.5}$\AA, and EW$_{\rm{H}_{\beta}} = 247.6^{+32.7}_{-46.1}$\AA.

Therefore we conclude that the Balmer line EWs further support our interpretation that the Recent Blue Monsters are dominated by young and very hot stellar populations, while the Blue Zombies have a non-negligible contribution from older stars ($\sim$400 Myr).

\subsection{Conclusions}\label{conclusion}

Our analysis of the BoRG-\jwst\ $z\sim8$ galaxies helps to reconstruct the puzzle of the $z\gtrsim10$ UV-luminous galaxies found with \jwst.  We have the unique opportunity of accessing the rest-optical emission of these galaxies to make predictions on the properties of the $z\gtrsim10$ blue monsters. Our main results can be summarized as follows:

\begin{enumerate}
\item The BoRG-\jwst\ galaxies are consistent with having negligible dust extinction, based on the $\beta_{\rm{UV}}$ slope and Balmer decrement.
\item We find no strong evidence for AGN activity based on the emission lines available, although some contribution from four sources cannot be ruled out, with one of them (1747\_817) being marginally consistent with an AGN in the OHNO diagnostic. Otherwise, both strong star formation and AGN contributions are consistent with the data. 
\item Half of the BoRG-\jwst\ galaxies have experienced an early burst of star formation at $z>12$, after which they appeared briefly as ``Blue Monsters", then faded, and then experienced a second major burst shortly before the time of observations. We call these galaxies ``Blue Zombies". The other half are caught shortly after their main burst of star formation, and are thus named ``Recent blue monsters."

\end{enumerate}

Given the similarity between BoRG-\jwst galaxies and ``blue monsters", we suggest that the same interpretation can be applied to the bright $z\gtrsim10$ population. In other words, they could be "Blue zombies" caught during their first major burst of star formation.    

An important factor that deserves further investigation is the role of environment. The BoRG-\jwst\ sample was selected from random pointing in the sky as part of a random pointing survey. At the moment, there is insufficient information to measure environmental parameters such as galaxy overdensity or halo mass. Precise environment measures for the BORG-\jwst sample and the ``blue monsters" are needed to make progress on this front.

%====================================================
%============ Acknowledgments ==========================
%====================================================
\acknowledgements
This work is based on observations made with the NASA/ESA/CSA James Webb Space Telescope. The data were obtained from the Mikulski Archive for Space Telescopes at the Space Telescope Science Institute, which is operated by the Association of Universities for Research in Astronomy, Inc., under NASA contract NAS 5-03127 for \jwst. 
The specific observations analyzed can be accessed via \dataset[doi: 10.17909/9s79-m291]{https://doi.org/10.17909/9s79-m291}. These observations are associated with programs \# 1747 and 2426. 
Funding from NASA through STScI awards JWST-GO-01747 and JWST-GO-02426 is gratefully acknowledged.

\facilities{\jwst}
\software{Astropy \citep{the_astropy_collaboration_astropy_2013},
          gsf \citep{morishita_massive_2019},
          Matplotlib \citep{hunter_matplotlib_2007},
          Numpy \citep{harris_array_2020},
          SciPy \citep{virtanen_scipy_2020},
          PyNeb \citep{luridiana_pyneb_2013}
      }
          
%=====================================================
%============ References ===========================
%=====================================================
\bibliographystyle{yahapj}
\bibliography{references.bib}      

\end{document}